\documentclass[a4paper,fleqn,usenatbib]{mnras}

\usepackage[T1]{fontenc}
\usepackage{ae,aecompl}
\usepackage{amsmath}
\usepackage{pbox}

\usepackage{graphicx}	
\usepackage{amsmath}	
\usepackage{amssymb}	
\usepackage{gensymb}
\usepackage{subfig}

\usepackage{todonotes}

\title[Disk winds in hard-state black-hole binaries]{Thermal and radiation driving can produce observable disk winds in hard-state X-ray binaries}

\author[N. Higginbottom et. al]
{Nick Higginbottom,$^{1}$\thanks{E-mail: nick\_higginbottom@fastmail.fm}
Christian Knigge$^{1}$, Stuart A Sim$^{2}$, Knox S. Long$^{3,4}$,
 \newauthor{James H. Matthews$^{5}$, Henrietta A. Hewitt$^{2}$, Edward J. Parkinson$^{1}$ and}
  \newauthor{Sam W. Mangham$^{1}$.}
\\
$^{1}$School of Physics and Astronomy, University of Southampton, Highfield, Southampton, SO17 1BJ, UK\\
$^{2}$School of Mathematics and Physics, Queen's University Belfast, University Road, Belfast, BT7 1NN, UK\\
$^{3}$Space Telescope Science Institute, 3700 San Martin Drive, Baltimore, MD, 21218, USA\\
$^{4}$Eureka Scientific Inc., 2542 Delmar Avenue, Suite 100, Oakland, CA, 94602-3017, USA\\
$^{5}$Institute of Astronomy, University of Cambridge, Madingley Road, Cambridge CB3 0HA, UK\\
}

\date{Accepted XXX. Received YYY; in original form ZZZ}

\pubyear{2019}

\hypersetup{draft}
\begin{document}
\label{firstpage}
\pagerange{\pageref{firstpage}--\pageref{lastpage}}
\maketitle

\begin{abstract}
X-ray signatures of outflowing gas have been detected in several accreting black-hole binaries, 
always in the soft state. 
A key question raised by these observations is whether these winds might also exist in the hard 
state. Here, we carry out the first full-frequency radiation hydrodynamic simulations of luminous 
($\rm{L = 0.5 \, L_{\mathrm{Edd}}}$) black-hole X-ray binary systems in both the hard and the 
soft state, with realistic spectral energy distributions (SEDs). Our simulations are designed 
to describe X-ray transients near the peak of their outburst, just before and after the hard-to-soft
state transition. At these luminosities, it is essential to include radiation driving, and we 
include not only electron scattering, but also photoelectric and line interactions. 
We find powerful outflows with $\rm{\dot{M}_{wind} \simeq 2 \,\dot{M}_{acc}}$ are driven by 
thermal and radiation pressure in both hard and soft states. The hard-state wind is significantly 
faster and carries approximately 20 times as much kinetic energy as the soft-state wind. However,
in the hard state the wind is more ionized, and so weaker X-ray absorption lines are seen over a 
narrower range of viewing angles. Nevertheless, for inclinations $\gtrsim 80^{\circ}$, blue-shifted 
wind-formed Fe~{\sc xxv} and Fe~{\sc xxvi} features should be observable even in the hard state. 
Given that the data required to detect these lines currently exist for only a single system in a 
{\em luminous} hard state -- the peculiar GRS~1915+105 -- we urge the acquisition of new observations
to test this prediction. The new generation of X-ray spectrometers should be able to resolve the velocity structure.

\end{abstract}

\begin{keywords}
Accretion discs -- hydrodynamics -- methods:numerical -- stars:winds -- X-rays:binaries
\end{keywords}



\section{Introduction}
\label{section:introduction}

Low mass X-ray binaries (LMXBs) are systems in which a low mass star loses mass onto a neutron
star or black hole companion, probably via Roche lobe overflow. 
This accretion takes place via an accretion disk.  In common with many other accreting systems, 
disk winds have been detected, albeit in a relatively small number of the total of about 200 such 
systems \citep{2007A&A...469..807L}.

Thermal driving is likely to be the wind driving mechanism  \citep{1983ApJ...271...70B,1996ApJ...461..767W,2002ApJ...565..455P} but
magnetic effects \citep{1992MNRAS.259..604T,1995ApJ...440..742H,2015MNRAS.454.1117S} may also be 
important and even dominate in some systems 
\citep[but also see \citealt{2006ApJ...652L.117N,2015MNRAS.451..475U,2016ApJ...823..159S}]
{2006Natur.441..953M,2008ApJ...680.1359M,2009ApJ...701..865K}. Thermal driving will be important if the 
geometry of the accretion disk and central source are favourable so that the surface of the accretion 
disk is illuminated by radiation from the central object/corona or inner parts of the disk. 
The disk atmosphere will then be strongly heated and expand away from the disk. If the thermal velocity of the 
gas exceeds the local escape velocity, a strong wind can be driven. This type of wind is most likely
to exist in systems with a large accretion disk (so outer parts of the disk have a relatively low escape velocity) 
and a high radiation temperature. The outflows observed to date in LMXBs tend to exist preferentially in systems
that appear to host large accretion disks \citep{2016AN....337..368D}.

 LMXBs are highly variable sources, 
 spending large periods of time in a low-luminosity hard SED (low-hard) state. Occasionally they brighten
 whilst maintaining a hard SED reaching luminosities of up to, or even exceeding, the Eddington luminosity
 for the central object mass ($\rm{L_{\mathrm{Edd}}}$). The sources then undergo a change of SED, switching to
 a softer, disk dominated (high-soft) state \citep{2010LNP...794...53B,2010MNRAS.403...61D}. 
 
 It is in this high-soft state that winds are detected in X-rays \citep{2012MNRAS.422L..11P}
 - hence the hypothesis that the wind is associated
 with the disk. The behaviour of XRBs in the high soft state is complex, and often many cycles of SED and 
 luminosity change are observed as the source returns over the course of weeks to the low-hard
  state and re-enters quiescence \citep[e.g.][]{2009MNRAS.400.1603M,2010A&A...522A..99C,2019MNRAS.482.1587B}. 
  The existence of a strong wind is a possible explanation for the 
  variability, and could also provide a mechanism for the transition from the soft, disk dominated
  state back to the hard state where the disk is truncated \citep[although see][]{2019arXiv190913601D}.
  
  Previously, we have carried out a series of simulations
  of thermally driven disk winds without radiative driving and we demonstrated that for plausible system parameters and
  SEDs, fairly strong winds are produced. \citet[][hereafter referred to as HK18]
  {2018MNRAS.479.3651H} showed that
  for reasonable system parameters based upon GRO-J1655-40 in an intermediate soft state
  with $\rm{L\sim 0.04~L_{\mathrm{Edd}}}$, the wind mass-loss rate was about twice the accretion rate
  required to generate the system luminosity. \citet[][hereafter HK19]{2019MNRAS.484.4635H} 
  further demonstrated
  that this mass-loss efficiency was approximately constant with source luminosity. 
  
  \cite{2018MNRAS.473..838D} included a simple treatment of 
  radiation driving and showed that as the source luminosity increased past $\rm{L=0.2~L_{\mathrm{Edd}}}$ the
  wind efficiency increases markedly. They also predicted a strong wind at low luminosities in the hard
  state, which is not detected due to over-ionization. 
  
  \cite[][hereafter TD19]{2019MNRAS.490.3098T} carried out simulations of H 1743-322 and included an improved
  treatment of radiation driving. They demonstrated the importance of including radiation driving, and suggested
  that not only is electron scattering important, but line driving and bound free interactions
  can be important for their fiducial simulation with $\rm{L=0.3~L_{\mathrm{Edd}}}$. They also investigated the effect
  of SED, as the source transitions from soft to hard at $\rm{L=0.06~L_{\mathrm{Edd}}}$. They demonstrated that in
  addition to a possible over-ionization effect, if an optically-thick ``failed-wind'' exists at 
  small radii, then this ``shadowing'' could block radiation from reaching the outer disk and therefore suppress the wind.
  
So, observations and simulations suggest that as a source transitions from the soft state to the hard
state at low luminosities, any thermal wind might be expected to disappear either due to a failure
in the driving mechanism through shadowing or through over-ionization. However, there is a gap in 
the story as a source brightens and then softens (although \emph{optical} outflows have been seen in
some sources in this portion of an outburst \citep{2016Natur.534...75M,2018MNRAS.479.3987M,2019MNRAS.489.3420J}.
We therefore wish to investigate whether we might expect to observe outflowing material in the hard
state at high luminosities - that is as the source is moving into outburst. To improve on our earlier efforts,
we have modified our method described previously to include both line and bound-free
  driving. We also use more realistic SEDs than previously, in order to properly separate the hard and 
  soft states.

\section{Method}
\label{section:method}

We use PLUTO (v4.3)\footnote{We previously used 
 \textsc{zeus2D} \citep[][extended by \citealt{2000ApJ...543..686P}]{1992ApJS...80..753S}. 
 Regression tests against our previous simulations show that the two codes give very 
 similar results.}
 hydrodynamics code
\citep{2007ApJS..170..228M} to solve the equations of hydrodynamics and 
{\sc python}\footnote{More information about and the source code for 
 \textsc{python} can be found at \url{https://github.com/agnwinds/python}.} \citep[][extended by 
  \citealt{2005MNRAS.363..615S}, \citealt{,2013MNRAS.436.1390H} and
 \citealt{2015MNRAS.450.3331M}]{2002ApJ...579..725L} to deal with radiative transfer.
 The two are coupled via an operator splitting formalism, with \textsc{python} 
 supplying heating and cooling rates along with radiation driving accelerations. 
 Our method is similar to that described by HK18 and HK19, except that we now explicitly 
include radiation driving. 
 
 Calls to \textsc{python} are made every 1000~s of hydrodynamic ``simulation time'' with the density, 
 temperature and velocity structure being passed between the codes. In \textsc{python},
 a population of Monte-Carlo radiation packets are generated to sample the requested SED, each 
 with weight W (ergs). The sum of all the packet weights is equal to the total luminosity
 of the source. These packets are
 then propagated through the simulation domain - all starting from the surface of a $7\times10^8$~cm
 sphere centred on the origin. This represents both 
 the X-ray emitting corona and any UV bright portions of the inner accretion disk. 
 The electron scattering ($\rm{g_{es}}$) and bound-free ($\rm{g_{bf}}$) accelerations are 
 estimated directly from the radiation packets in the radiation transfer calculation as
 \begin{equation}
 \overline{g}_{es}=\frac{n_e}{c\rho V}\sum_{paths}\sigma(\nu)_{KN}W_{ave}\delta\overline{s}~cm~s^{-2},
 \end{equation}
and
\begin{equation}
\overline{g}_{bf}=\frac{1}{c\rho V}\sum_{paths}\delta\overline{W}~cm~s^{-2}.
\end{equation}
Here, $W_{ave}$ is the mean weight of a Monte-Carlo radiation packet 
along path segment $\delta s$ and
$\delta W$ is the weight reduction due to bound free opacity. V is the volume of the
cell in question, $\sigma(\nu)_{KN}$ is the Klein-Nishina cross section for 
electron scattering, $n_e$ is the electron number density and $\rho$ is the mass density, 

 The line-driving acceleration is calculated using the method described by 
 \cite{2013ApJ...767..114P}. This closely follows the classic line force
 multiplier method first described by \cite{1975ApJ...195..157C}, 
 where the radiative acceleration is given by 
 \begin{equation}
     \overline{g}_{rad}=\frac{n_e \kappa_e}{\rho c}\overline{F} M(t).
     \label{equation:g}
 \end{equation}
In this equation, $\kappa_e$ is the electron scattering
 opacity, and $F$ is the radiative flux. The quantity $M(t)$ is the so-called 
 ``force multiplier'' -- the ratio between the accelerations due to line-driving 
 and electron-scattering -- and 
 depends on the dimensionless optical depth parameter, $t$ (see Equations~\ref{mult}
 and \ref{opt} below). 
 
For the purpose of calculating $M(t)$, we split the flux into three bands: 
visible (up to $7.4\times10^{14}$ Hz), 
ultraviolet (UV -   between $7.4\times10^{14}$ and $3\times10^{16}$ Hz) and 
X-ray (above $3\times10^{16}$ Hz). 
This approach allows a more accurate calculation of $M(t)$ and also
allows the direction of the acceleration vector in each band to be different; the net 
 acceleration is then simply the vector sum over all three bands. This flexibility allows us to take 
 account of complex illumination geometry which is important for line driven winds in CVs and AGN where
 the wind probably arises above the UV bright portion of the disk \citep{2018MNRAS.481.2745D}.

 The radiative flux in each band is calculated from the radiation packets as
 
 \begin{equation}
\overline{F}=\frac{1}{V}\sum_{paths}w_{ave}\delta\overline{s},
\label{equation:f_bar}
\end{equation}
which implicitly takes account of any frequency dependant attenuation between the source
and the cell in question. This represents a significant advance over previous calculations that adopted a grey opacity (i.e. the attenuation is assumed to be 
frequency \emph{in}dependant). 
 
 The force multiplier is similarly banded and, in each band, is estimated as
 \begin{equation}
 \label{mult}
M(t)=\sum_{lines}\Delta\nu_D\frac{j_{\nu}}{j}\frac{1-\exp(-\eta_{u,l}t)}{t}
\end{equation}
Here, $\Delta\nu_D$ is the Doppler width of each line, $j$ is the frequency-averaged 
mean intensity, and $\eta_{u,l}$ is the ratio of populations in the upper and lower 
levels linked by the line transition. The specific mean intensity, $j_{\nu}$, in each band is calculated
from a spectral model generated by energy packets in the cell and therefore also takes account of frequency dependent attenuation. 

\begin{figure}
\includegraphics[width=\columnwidth]{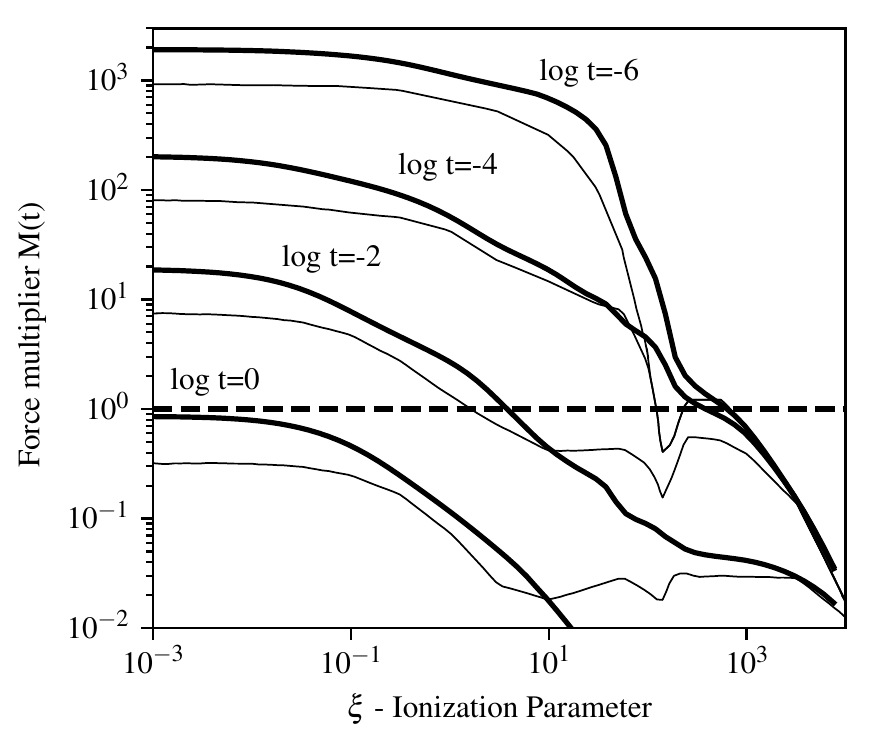}
\caption{The force multiplier for line driving as a function of ionization
parameter $\rm{\xi}$ for four values of the optical depth parameter, t (heavy
lines). Fine lines are from \citet{1990ApJ...365..321S} for comparison. The
horizontal dashed line shows the level at which line driving becomes sub-dominant.}
\label{figure:M_vs_xi}
\end{figure}

The optical depth parameter $t$ is given by
\begin{equation}
\label{opt}
    t=\kappa_e\rho v_{th} \left(\frac{dv}{dr}\right)^{-1},
\end{equation}
 where $v_{th}$ is the thermal velocity, and the term in brackets is the velocity
 gradient of the plasma in the direction of the local radiation flux. Figure \ref{figure:M_vs_xi}
 shows the values of $M(t)$ obtained with this method in a setting that can be directly compared
 to Figure~2 of \cite{1990ApJ...365..321S}. Specifically, the medium considered here is 
 an optically thin photo-ionized cloud, whose 
 ionization state and temperature are controlled by an irradiating 10~keV bremsstrahlung spectrum. The 
 line-driving flux ($\overline{F}$; Equation~\ref{equation:f_bar}) is taken to be a Kurucz stellar
 atmosphere with $\rm{T_{eff} = 25000~}$K and $\rm{\log{g}=3}$ \citep{1993ASPC...44...87K}.
 M(t) is plotted against the ionization
 parameter $\rm{\xi}$, defined as $\rm{\xi=L_x/(n_H r^2})$ where $\rm{L_x}$ is the
 luminosity above 13.6eV and r is the distance between the source of radiation and the cloud.

 Agreement with 
 \cite{1990ApJ...365..321S} is within a factor of 2 for values of $\rm{\xi}$ where 
$\rm{M > 1}$. For high values of $\rm{\xi}$, the larger disagreement for high values of t is probably 
due to differences in the line lists, particularly 
highly ionized ions of Iron. However, in such cases radiative driving would be 
 dominated by electron scattering - there is effectively a floor of $M=1$, below which
 line driving is sub-dominant.

As in HK19, we use a logarithmic grid in both the $r$ and $\theta$
directions. The $\theta$ grid has 100 cells, running from 0\degree to 
90\degree, whilst the $r$ grid has 220 cells running from 
$6.2\times10^9$~cm (1/20 of the Compton radius for the hard SED - 
6000 gravitational radii) to
$10^{13}$~cm. 
Observations suggest that the inner edge of the accretion disk is truncated at
100s to 1000s of gravitational radii in the 
hard state \citep{2015A&A...573A.120P}, so our inner edge is at
a radius where a disk is expected to exist in both the soft and hard states.
The outer edge of the disk is set to a radius 
of $10^{12}$cm. We set a midplane density boundary condition 
such that $\rm{\rho(r)=\rho_0(r/r_0)^{-2}}$ where $\rm{r_0=5\times10^{11}}$cm.
This acts as a mass reservoir for the outflow.

\begin{figure}
\includegraphics[width=\columnwidth]{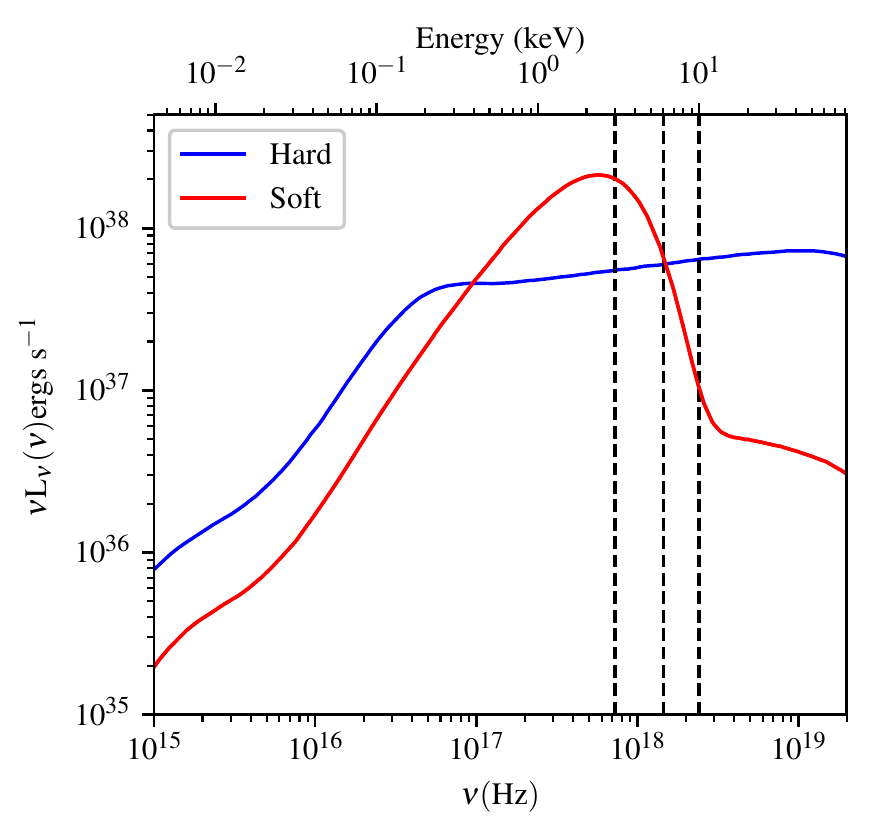}
\includegraphics[width=\columnwidth]{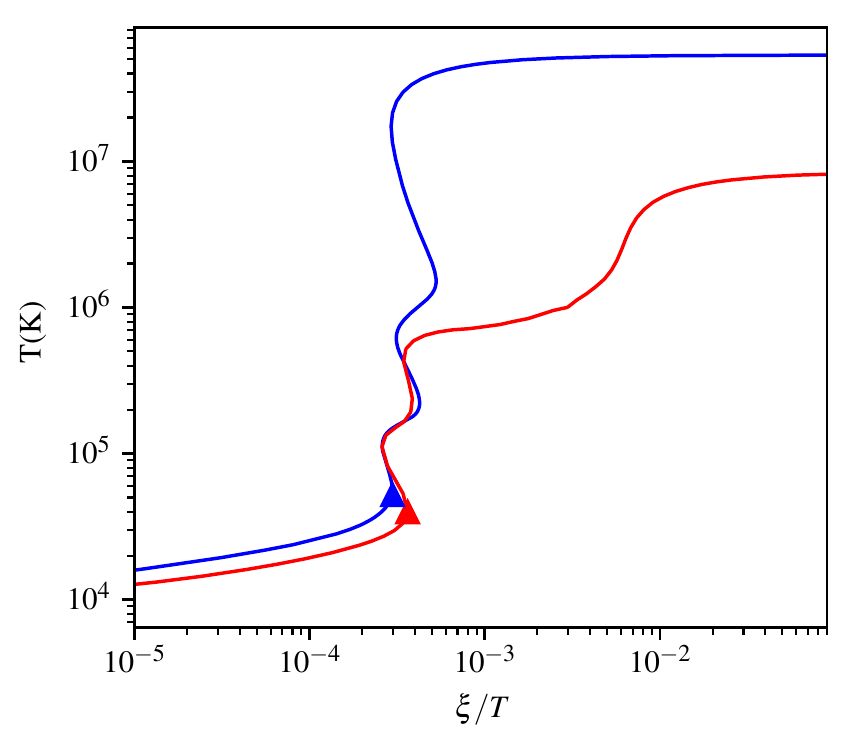}

\caption{SEDs used in the simulations (upper panel) and the stability
curves for the same two SEDs (lower panel); the colour scheme is the same
for both plots. The vertical dotted lines show the band limits
used to define the hardness ratios which are
hard state: $\log(L_{6-10\rm{keV}}/L_{3-6\rm{keV}})=-0.1$,
soft state: $\log(L_{6-10\rm{keV}}/L_{3-6\rm{keV}})=-0.8$. The symbols on the 
lower plot
mark the location of $\rm{\xi_{cold,max}}$, the point at which the gas first 
becomes thermally unstable.}
\label{figure:SED}
\end{figure}

Our method of calculating heating and cooling rates -- as well as 
radiation driving -- with our RT code allows us to use realistic 
SEDs. Here, we use hard and soft SEDs based upon those
reported by \cite{2018MNRAS.473..838D}, which are in turn based on   
observations of H1743-322. More specifically, we adopt
their $\rm{0.5~L_{\mathrm{Edd}}}$ SED for the soft state, and their $\rm{0.01~L_
{\mathrm{Edd}}}$ SED for the hard state, but scaled again to  $\rm{0.5~L_{\mathrm{Edd}}}$. 
Figure~\ref{figure:SED} shows the SEDs we use and the corresponding stability 
curves in the optically thin limit.

\begin{table} 
\begin{tabular}{p{4.0cm}p{.5cm}p{.5cm}p{.6cm}p{.6cm}} 
\hline  
& \multicolumn{4}{c}{Physical Parameters}\\ 
\hline  
SED  &\multicolumn{2}{c}{Soft} & \multicolumn{2}{c}{Hard} \\ 
\hline 
$\rm{L_x~(10^{37}}~\rm{ergs~s^{-1}})$ & \multicolumn{4}{c}{44.1 (0.5 $\rm{L_{Edd}}$)} \\ 
$\dot{\rm{M}}_{\rm{acc}}~\rm{g~s^{-1}})$& \multicolumn{4}{c}{$5.9\times10^{18}$} \\
$\rm{\dot{M}_{acc}~(M_{\odot}yr^{-1})}$& \multicolumn{4}{c}{$6.8\times10^{-8}$} \\ 
$\rm{M_{BH}~(M_{\odot})}$  & \multicolumn{4}{c}{7} \\ 
$\rm{T_{IC}~(10^6~}$K)   & \multicolumn{2}{c}{8.46} & \multicolumn{2}{c}{53.5} \\ 
$\rm{\xi_{cold,max}}$ & \multicolumn{2}{c}{14.4} & \multicolumn{2}{c}{15.5} \\ 
$\rm{R_{IC}~(10^{11}~cm)}$  & \multicolumn{2}{c}{  7.9} & \multicolumn{2}{c}{  1.2} \\ 
$\rm{\rho_0~(10^{-10}~g~cm^{-3})}$ & \multicolumn{2}{c}{  3.0} & \multicolumn{2}{c}{  2.8} \\ 
\hline
& \multicolumn{4}{c}{Derived wind properties}\\ 
\hline 
Radiation driving? & N & Y & N & Y \\ 
\hline 
$\rm{V_r(max,\theta>60\degree,~{km~s^{-1}})}^1$ & 470 & 510 & 1270 & 1560 \\ 
$\rm{T(max,\theta>60\degree,r>R_{IC},~10^6K})^1$ &   8.4 &   7.9 &  47.7 &  47.1  \\ 
$\rm{V_{th}(max,\theta>60\degree,~{km~s^{-1}})}^1$ & 340 & 330 & 810 & 800  \\ 
$\rm{N_H~(70\degree)~(10^{22}}~\rm{cm^{-2})}$  &   9.9 &  11.4 &  21.8 &  17.8  \\ 
$\rm{N_H~(80\degree)~(10^{22}}~\rm{cm^{-2})}$  &  32.6 &  35.3 &  37.8 &  27.5 \\ 
${\dot{\rm{M}}_{\rm{wind}}~(10^{18}~\rm{g~s^{-1}}})$  &   7.7 &   9.7 &  10.4 &  12.6  \\ 
${\dot{\rm{M}}_{\rm{wind}}/\dot{\rm{M}}_{\rm{acc}}}$  &   1.3 &   1.6 &   1.8 &   2.1  \\ 
${0.5\dot{\rm{M}}\rm{V_r^2}~(10^{34}}~\rm{erg~s^{-1}})$ &   1.0 &   1.5 &  13.0 &  19.5  \\ 
Angle for EW(Fe \textsc{xxv})>5eV                 & -      & 70\degree & - & -  \\ 
Angle for EW(Fe \textsc{xxvi})>5eV               & -         & 60\degree  & - & 75\degree \\

\hline
\end{tabular}
\caption{Parameters adopted in the simulations, along with key properties of the resulting outflows. 
(1 - these properties are given as maximum values for the parts of the wind
where absorption features tend to be seen - that is angles above about 60\degree)}
\label{table:wind_param}
\end{table}

\section{Results}
\label{section:results}

\subsection{Physical Outflow Properties}

\begin{figure*}
\includegraphics[width=\columnwidth]{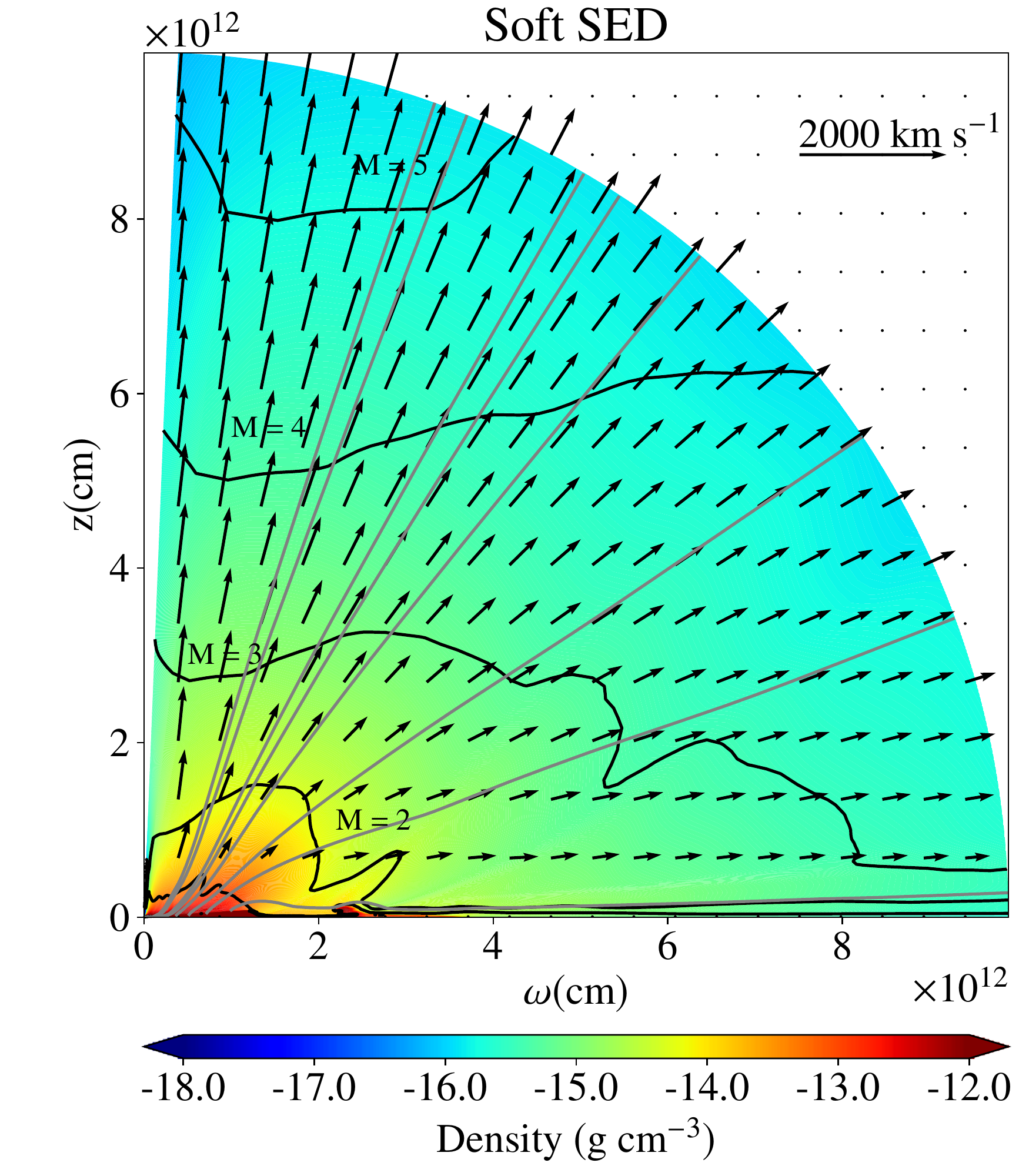}
\includegraphics[width=\columnwidth]{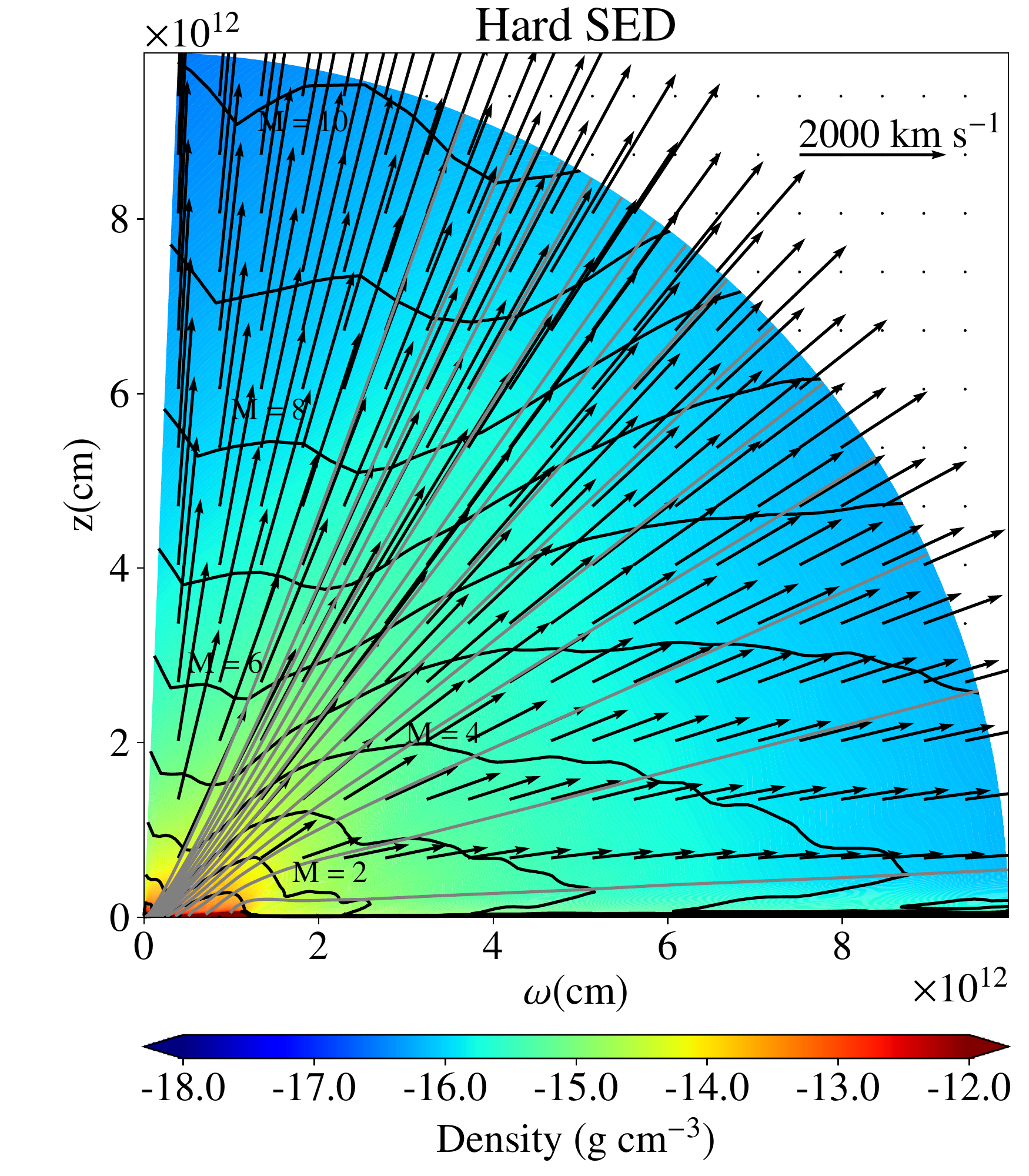}
\caption{The density (colours) and velocity structure (arrows) of the stable final states for 
the hard and soft SED simulations, with all radiation driving mechanisms enabled. 
Grey lines show streamlines, and the black line shows the location of the Mach surfaces.}
\label{figure:wind_small_image}
\end{figure*}

The hard- [and soft-] state simulations reach stable states after about $5\times10^5$~s [$10\times10^5$~s], which corresponds to 
roughly 5 sound crossing times.
The final snapshots of the simulations are shown in 
Figure \ref{figure:wind_small_image}. To first order, the density structure is very similar to comparable cases given by \cite{2019MNRAS.490.3098T} over matching radial range (inside about $7\times10^{11}$~cm).
Some key physical properties of the outflows are summarized in the 
lower section of Table \ref{table:wind_param}. 

A hydro-statically supported structure develops above the midplane in 
both simulations. This is convex with a half opening angle of 3/4\degree~ in the hard/soft states. This can be thought of as a disk atmosphere, overlying any
dense disk that exists in the system. Above that structure,
both hard-state and soft-state simulations produce powerful outflows with comparable efficiencies,  ${\dot{\rm{M}}_{\rm{wind}}/\dot{\rm{M}}_{\rm{acc}}} \simeq 2$. These winds represent important mass and 
energy sinks for XRBs, although not to the degree required for destabilizing the disk \citep[c.f.][but also see \citealt{2020arXiv200103791G}]{1986ApJ...306...90S}.
The outflows are thermally driven, but assisted by radiation pressure. Based on test runs in which radiation forces have been artificially switched off, radiation pressure acts mainly to increase the average radial velocity (by $\simeq$20\% in the hard state and by $\simeq$50\% in the soft state) and slightly to increase the mass-loss rate (by $\simeq$20\% in both cases). Characteristic values of the force multiplier in the acceleration zone of the outflows are only very slightly over 1 in the hard state and $M \simeq 1.2$ in the soft state, indicating that this radiation pressure is mainly exerted on free electrons. The modest increase in $\dot{M}_{wind}$ can be ascribed to the larger pressure drop above the disk in the faster, radiation-assisted wind. This promotes evaporation in the upper layers of the disk atmosphere.

The most obvious difference between the hard-state and soft-state simulations is that considerably faster wind speeds are 
attained in the hard state, although because the density at the outer radius is lower in the hard state the massloss rates are similar. This is not the result of radiation driving, since the accelerations due to the radiation field
are very similar in both simulations. Instead, the different speeds are due to the distinct stability curves
produced by the two SEDs \citep[see][for an investigation of this phenomenon]{2017MNRAS.467.4161D}. As shown in Figure
\ref{figure:SED}, the soft state 
curve has an intermediate stable temperature at about $1\times10^6$K, whereas the hard state curve 
is effectively unstable all the way up to the Compton temperature (which is much higher than for the soft-state). As a result, gas heating and expansion are more explosive in the hard state, and gas at the higher Compton temperature has a faster thermal speed. These two factors give rise to a faster outflow. The combination of higher wind speeds and 
similar mass-loss rates also means that $\simeq 20$ times more kinetic energy is carried away by the hard-state outflow.

\subsection{Observable Outflow Signatures}

As described in Section~\ref{section:method}, we use our stand-alone, multi-frequency Monte Carlo ionization and radiative transfer code \textsc{python} as the radiation module in our RHD simulations. One of the benefits of this arrangement is that we can easily generate synthetic spectra for any desired line of sight. These spectra take into account 
the velocity, density and ionization structure of the wind. Critically, they also include the
effects of multiple scattering. In order to generate these detailed spectra, we take the final, approximately steady states of the simulations and generate a new population of photon packets over a restricted frequency band of interest. 
We then propagate these packets though the domain and extract the spectra they would produce for different observer orientations.\footnote{for a more detailed discussion of this part of the code see \cite{2002ApJ...579..725L}.}

\begin{figure*}
\includegraphics{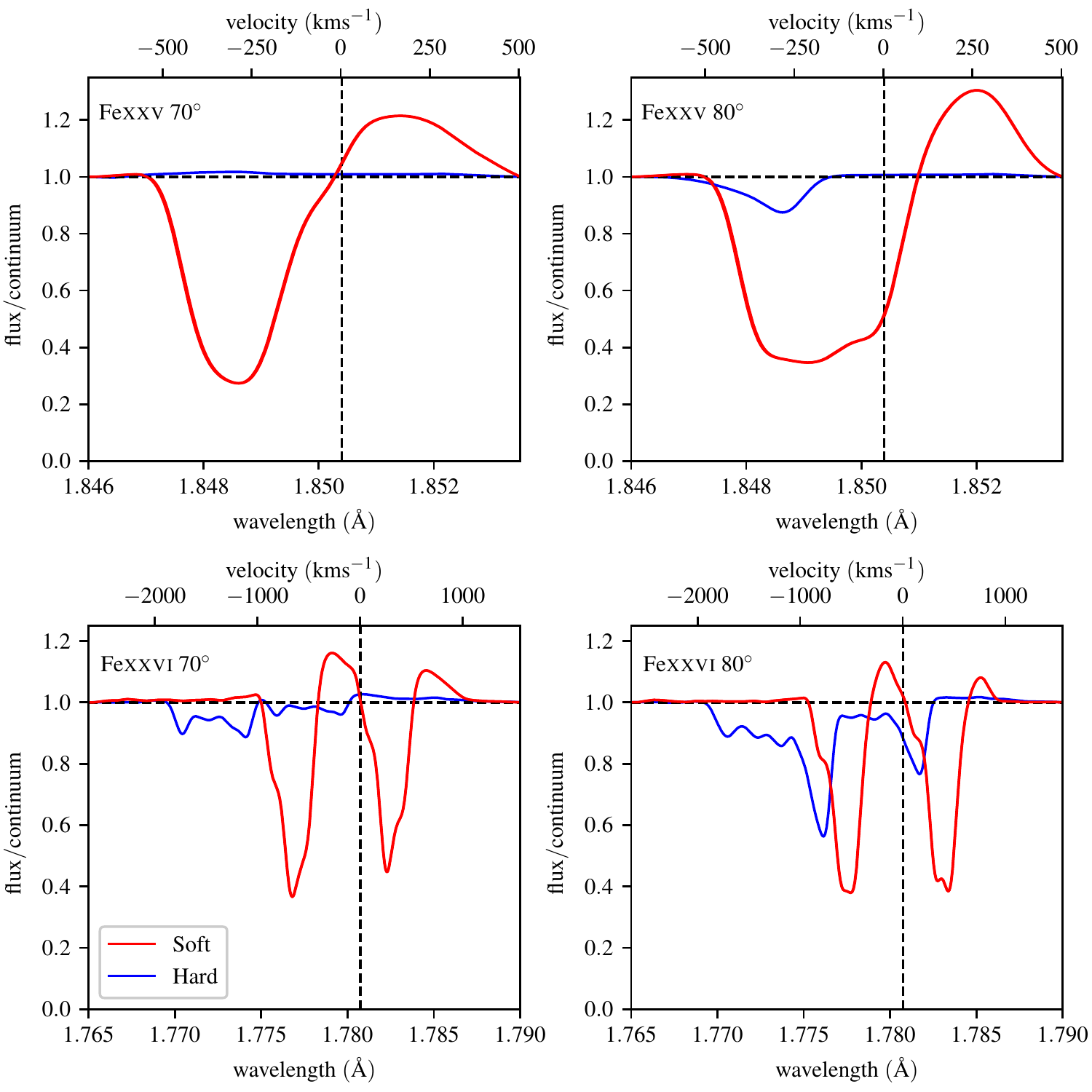}
\caption{Monte-Carlo spectra showing the absorption and scattering profiles
for the Fe~\textsc{xxv} 
(1.85~{\AA}-upper) and Fe~\textsc{xxvi} K$\rm{\alpha}$ (1.78~{\AA}-lower) at $70\degree$ 
(left) and $80\degree$ (right)}
\label{figure:fe25_80}
\end{figure*}

Figure \ref{figure:fe25_80} shows the spectra synthesized in this way for two sightlines \footnote{As mentioned earlier, our simulation domain starts at the surface of any underlying accretion disk, and so the angles for the sightlines presented here are relative rather than absolute. 80\degree~ is therefore 10\degree~ above the surface of the disk}, focusing on the ground state
transitions of Fe \textsc{xxv} and Fe \textsc{xxvi}. The raw spectra have been smoothed with a Gaussian of FWHM of 2.5~eV  - the expected energy resolution of the X-IFU spectrometer to be flown on the Athena X-ray telescope \citep{2014cosp...40E2556P}. The vertical dashed 
lines show the rest wavelength of the Fe \textsc{xxv} line, and the mean wavelength of the Fe \textsc{xxvi} doublet. We have also 
generated optical spectra, to see if we observe the optical lines seen in some XRBs \citep{2016Natur.534...75M,2018MNRAS.479.3987M,2019MNRAS.489.3420J}. We see no features here, because the wind is simply 
far too ionized. However, since X-ray and optical features have not been observed at the same time to date, there is no reason
to assume that the same simulation \emph{would} produce the correct conditions for both.

The difference in wind velocity between the hard- and soft-state winds are apparent in these spectra. The 
soft-state line profiles exhibit blue-shifted absorption out to velocities of about 400~$\rm{km~s^{-1}}$, whereas the hard-state profiles show weaker absorption features, but out to more than 1000~$\rm{km~s^{-1}}$.

\begin{figure}
\includegraphics[width=\columnwidth]{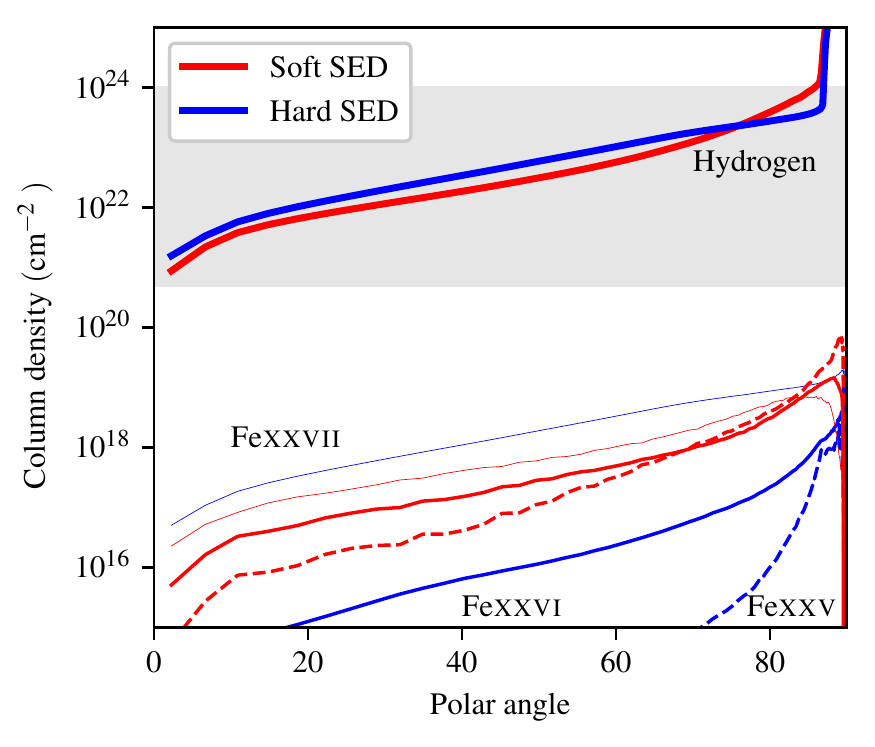}
\caption{The column density of Hydrogen (heavy lines), Fe\textsc{xxv} (broken lines), Fe\textsc{xxvi} (solid, thin lines) and Fe\textsc{xxvii} (very faint lines)
 for the soft and hard state simulations. The grey region shows the range of  column densities for XRB winds inferred from observations.}
\label{figure:column}
\end{figure}

We can understand these differences by examining the column density of the wind in the different ions. Figure
\ref{figure:column} shows this and we firstly see that the total Hydrogen column density of the wind in both states is similar. The wind is optically thin until the dense
accretion disk region is reached in both cases and the values fit in the observed range for LMXBs ($0.5\times10^{21}\rightarrow10^{24}~\rm{cm^{-2}}$; \citealt{2016AN....337..368D}), shown as the shaded area on Figure \ref{figure:column}. However, the ionization state is very different. In the hard state, Iron in the the plasma is overwhelmingly fully ionized, with Helium-like Iron very rare until the dense midplane region is reached. In contrast, in the soft state the Iron column densities in the three ionization states shown are within an order of magnitude of each other - showing that Iron is far less ionized in the soft state and thereby explaining the significantly higher absorption.

Finally, it is interesting that we see not just blue-shifted absorption, but classic P-Cygni line profiles in the soft state. That is, broad, blue-shifted absorption accompanied by red-shifted emission. Such features have, in fact, been observed 
in XRBs \cite[e.g.][]{2000ApJ...544L.123B,2016ApJ...821L...9M} and also in ultra-fast outflows from AGN \citep{2015Sci...347..860N}. 
In our simulations, the red-shifted emission
feature is associated with photons that have been scattered {\em into} the line of sight by the wind. Such scattering features tend to be more or less symmetric about the rest wavelength  \cite[e.g.][]{1995MNRAS.273..225K,2002ApJ...579..725L}, 
so the net blue-shifted absorption trough is actually partly filled in by this scattered radiation.

\begin{figure*}
\includegraphics[width=\columnwidth]{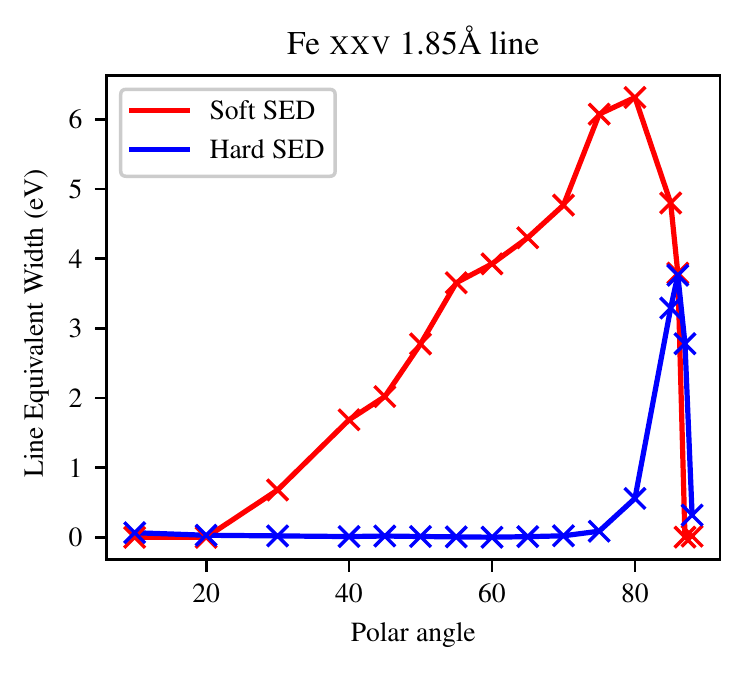}
\includegraphics[width=\columnwidth]{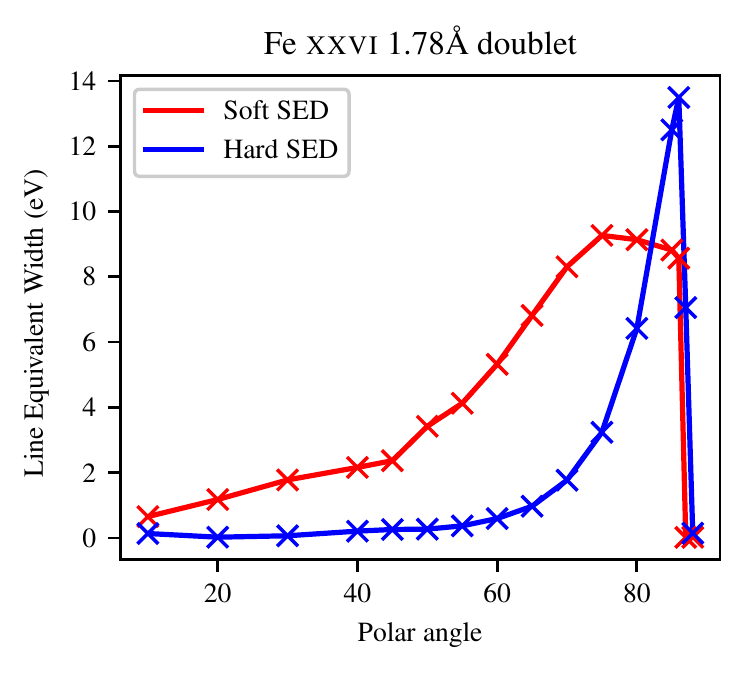}
\caption{The orientation dependence of the equivalent widths of the Fe~\textsc{xxv} 
(1.85~{\AA}) and Fe~\textsc{xxvi} K$\rm{\alpha}$ (1.78~{\AA}) absorption
features in our simulations.}
\label{figure:ew_vs_theta}
\end{figure*}

To highlight the range of inclinations over which these winds could be observed, Figure~\ref{figure:ew_vs_theta} shows the absorption equivalent widths (EWs) of the same lines as a function of viewing angle. With current instrumentation, a reasonable lower limit on detectable Fe \textsc{xxvi} absorption is probably $\simeq 5$~eV \citep[c.f.][]{2012MNRAS.422L..11P}. Adopting this limit, Figure~\ref{figure:ew_vs_theta} suggests that, in the soft state, winds might be detectable for $i \gtrsim 60^{\degree}$. The EWs in Figure~\ref{figure:ew_vs_theta} are rather lower than seen in several observations; typically 30eV for Fe\textsc{xxv} and 50eV for Fe\textsc{xxvi} \citep[e.g.][]{2006ApJ...646..394M,2009ApJ...695..888U,2014A&A...571A..76D,2018ApJ...861...26A,2019MNRAS.482.2597G}. There are two most likely possible reasons for this underestimate. Firstly, the presence of a P-Cygni type profile reveals significant scattering from this transition from other sightlines, filling in the absorption trough. Since this is rarely seen in observations, and we are confident that our treatment of scattering captures the relevant physics,  it suggests that our solution has an excess of the relevant ions in the parts of the wind towards the pole. To check this we have calculated the EW  using the ``absorption only'' line-transfer mode in \textsc{python}, thereby removing the scattered component. The soft state EW is about 50\% larger for both lines.

Secondly, since both lines are very deep when scattered light is removed, the EW is sensitive to the velocity of the scattering material. If the wind were slightly faster, then the EW would increase. We know that our velocity is somewhat less than observed in many systems, and so this would also lead to an underestimate for the soft state EWs. Since the intrinsic absorption profiles are already nearly saturated, an increase in wind density, or an increase in the proportion of the wind in the relevant ionic species would not increase the EW. In any case Figure \ref{figure:column} shows that our overall column density is already similar to that inferred from observations and there is already a significant column density of Fe\textsc{xxv} and Fe\textsc{xxvi} in the soft state, relative to the total Iron column.

By contrast, hard-state winds would be detectable only for $i \gtrsim 75^{\degree}$. On the one hand, this result is consistent with X-ray wind features appearing preferentially in soft state XRBs. On the other hand, our simulations suggest that winds may be present in hard states also, and that they should be detectable in systems viewed close to edge-on. There is little scattered light in the hard state, and the lines are not saturated. Therefore the two reasons for underestimates of the soft state line EWs do not apply in the hard state. However, clearly a decrease in the ionization state could rapidly increase the visibility of both lines.

\section{Discussion}
\label{section:discussion}

\subsection{Comparison to other modelling efforts}

To the best of our knowledge, the XRB disk wind simulations presented here are the first RHD calculations with a self-consistent treatment of ionization, frequency dependant radiative transfer and radiation driving. However, two other types of RHD calculations {\em have} been carried out before, and it is worth briefly comparing these to our own.

First, the simulations presented in HK18 and HK19 used the same method that we use here
to calculate the local heating and cooling rates. That is, these rates include radiative heating and cooling terms that account for the wavelength-dependent opacity between the central X-ray source and any cell in the simulation domain. However, our earlier simulations did {\em not} include the pressure exerted directly by the radiation field on the gas. 
The system modelled in HK18 had a luminosity of only $\rm{L = 0.04~L_{\mathrm{Edd}}}$, so radiation forces could be safely neglected. However, in HK19, we explored the luminosity dependence of thermally driven disk winds, finding that ${\dot{\rm{M}}_{\rm{wind}}/\dot{\rm{M}}_{\rm{acc}}} \simeq 2$, independent of luminosity. This is similar to the outflow efficiencies in the simulations presented here. 

Given that the calculations in HK19 did not include radiation driving, one might actually have 
expected the inclusion of this process to increase the wind efficiency. However, as noted above, the main effect of radiation pressure is to increase the wind {\em speed} -- i.e. the radiation force primarily accelerates material that is already on its way out of the system. The comparison to HK19 is additionally complicated by the different SEDs used in the two studies. Specifically, HK19 used a bremsstrahlung SED, whereas here we adopted the more realistic SED shapes suggested by
\cite{2018MNRAS.473..838D}. This difference matters, because the shape of the stability curve, and hence the efficiency of the wind launching, depend strongly on the SED shape.

Second, TD19 carried out RHD simulations in which {\sc cloudy} \citep{2017RMxAA..53..385F} is used to determine the thermal and ionization state of the gas. Their calculations do account for both thermal and radiation driving, with the latter including all of the relevant contributions (i.e. radiation pressure associated with free-free, bound-free and bound-bound interactions). From the computational physics perspective, the main difference between their simulations and ours concerns the {\em shape} of the local SED in each cell. The calculations in TD19 do account for attenuation, but only in the gray limit. That is, each grid cell is illuminated by the same SED. In our simulations, the effect of frequency-dependent attenuation on the local SED is calculated self-consistently. As noted above, this difference matters, because the stability properties of the irradiated gas depend strongly on the SED shape.

However, these computational issues are probably less important than a key difference in the sub-grid physics adopted by us vs that adopted in TD19. Specifically, TD19 allow for attenuation of the X-ray flux hitting the disk by an optically thick atmosphere, or failed wind above the inner disk \footnote{This object is also referred to as a corona in TD19 - however it is distinct from the X-ray emitting corona surrounding the central object.} close to the central object. This blocking layer is not itself part of the computational domain, but TD19 estimate its properties based on the analytic model of \cite{1983ApJ...271...70B}. It shadows much of the outer disk, especially in the hard state. It is this preferential shadowing that inhibits the launching of a thermally driven wind in their hard state simulations. We have not included such an attenuating inner disk atmosphere in our simulations, and so an outflow {\em is} launched even in the hard state.

It is difficult to assess which of these prescriptions is closer to reality. The inner disk atmosphere is expected to 
arise within $\simeq 200~R_G$, and it is not even clear if the {\em disk} extends down to such radii in the hard state \citep{2015A&A...573A.120P,2018ApJ...855...61W}. Moreover, in our multi-dimensional RT calculations, such blocking structures are often less attenuating than one might expect, due to multiple scattering effects \citep{2010MNRAS.408.1396S,2014ApJ...789...19H}. Examining the contributions to the radiation field  in the simulations reported here, we find that scattered radiation is a significant contributor to the radiation field in the base of the acceleration zone where the gas starts to be heated. Nevertheless, the amount, ionization state and spatial distribution of any material at small radii is a major uncertainty for simulations focused on larger scales. If the hard-state wind signatures predicted by our simulations are not detected in sensitive observations of high-inclination systems, the type of shadowing envisaged by TD19 could well be responsible. However,  since the absorbing ions are already a minority species, a slight
increase in the ionization state of the wind could also cause the absorption features to vanish entirely.

\subsection{Comparison to observations}

For the soft state, the line profile shapes (Figure~\ref{figure:fe25_80}) and equivalent width distributions (Figure~\ref{figure:ew_vs_theta}) produced by our RHD simulation agrees quite well with observations. In particular, 
blue edge velocities of $\simeq 500~{\rm km~s^{-1}}$ and $\simeq 1000~{\rm km~s^{-1}}$ are seen in the 
Fe~\textsc{xxv} and Fe~\textsc{xxvi} transitions, respectively, for $i = 80\degree$. As found previously by TD19, the 
boost in wind speed provided by radiation driving helps significantly to bring theory in line with observations in this case.

However, our hard-state simulation also produces detectable wind signatures, at least in Fe \textsc{xxvi} transition, where the blue edge now reaches $\simeq 2000~{\rm km~s^{-1}}$ for $i = 80\degree$. As noted in Section~\ref{section:introduction}, no X-ray wind signatures have so far been found in any hard-state XRB. Does this mean our simulations are already in conflict with observations? 

We believe the answer is ``no'' -- or at least ``not yet''. Among all of the observations collected by \cite{2012MNRAS.422L..11P} and shown in their Figure~1, sensitive upper limits on the EWs of wind-formed Fe \textsc{xxvi} lines exist for only a single {\em luminous} ($L \gtrsim 0.2~L_{\mathrm{Edd}}$) hard-state system. Moreover, that system is GRS~1915+105, which is arguably one of the most idiosyncratic BH XRBs and does not really exhibit canonical hard and soft states \citep{2000A&A...355..271B}. In this system, \cite{2009Natur.458..481N} report an upper limit on the Fe \textsc{xxvi} line of 1 eV but note that a significant narrow absorption line with EW $\sim5$ eV had been previously reported in one of the observations \citep{2002ApJ...567.1102L}.

We are not aware of any additional upper limits that have become available since then for any other luminous hard-state systems. For example, \cite{2012ApJ...759L...6M}
reported strong upper limits of 2.1~eV and 1.6~eV on Fe~\textsc{xxv} and Fe~\textsc{xxvi} wind features, respectively, for H1743-322 in the hard state. However, the luminosity of the system was only $\rm{L \simeq 0.02~L_{\mathrm{Edd}}}$ at the time of these observations. To test whether the predicted hard-state winds should be detectable at these lower luminosities also, we plan to carry out a suite of hard-state simulations covering a wide range of luminosities (mirroring the set of simulations we presented in HK19 for the soft state). However, it is already clear that additional, high-quality data will be needed to either detect hard-state X-ray winds or rule out their presence. This will require time-critical observing campaigns targeting transient XRBs near the peak of their outburst, just before the hard-to-soft state transition.

There have also been several observations of systems in states which do not fit into clear hard/soft SED classifications. They are either transitioning from hard to soft states \citep{2012ApJ...750...27N}, soft to hard states \citep{2019MNRAS.482.2597G}  or increasing luminosity in a soft state \cite{2014A&A...571A..76D}. We have shown here that the SED strongly affects the structure (and ionization state) of the wind and hence how observable it is, and so these cases are also clear candidates for future simulations.

\section{Summary}
\label{section:summary}
We have carried out full-frequency radiation-hydrodynamic simulations of thermally and radiation-driven 
disk winds in luminous X-ray binary systems. Our model system is characterized by a luminosity of $L = 0.5~L_{\mathrm{Edd}}$, and we carry out calculations for two distinct spectral states (hard and soft), for which we adopt realistic spectral energy distributions. Ionization and radiative transfer are treated in detail in our simulations, and we account for the dynamical  effects of radiation pressure on both free and bound electrons. We also present synthetic line profiles and equivalent width distributions for the X-ray transitions in which wind signatures have been observed.

Our main result is that powerful and detectable outflows are produced in both simulations, although the X-ray wind lines are weaker -- and detectable over a smaller range of inclinations -- in the hard-state calculation. The mass-loss efficiency is comparable in both states --  $\dot{M}_{wind} \simeq 2 \,\dot{M}_{acc}$ -- but the hard-state wind is faster and carries away 
20 times as much kinetic energy as the soft-state wind. 

Comparing our results to the existing observational constraints, we find good agreement (at least qualitatively) for the soft state. This agreement is partly due to the inclusion of radiation pressure, which boosts the wind speed in the simulations closer to the observed levels.

No hard-state winds have been detected in X-ray observations to date, but we note that meaningful constraints currently exist for only a single {\em luminous} hard-state XRB. That system is GRS~1915+105, whose spectral states are highly unusual. We therefore urge the acquisition of new, sensitive X-ray spectroscopy of luminous, high inclination XRB in the hard state. This will require target-of-opportunity observations of transient XRBs, triggered after the rise to outburst but before the hard-to-soft state transition. Additionally, the increased spectral resolution of up-coming spectrometers such as X-IFU on Athena \citep[2.5eV][]{2014cosp...40E2556P} and XRISM \citep[5-7eV][]{2018arXiv180706903G} will permit the velocity structure of absorption features to be resolved.

\section{acknowledgements}
Calculations in this work made use of the Iridis5 Supercomputer at the
University of Southampton. NSH and CK acknowledge support by the
Science and Technology Facilities Council grant ST/M001326/1.  
KSL acknowledges the support of NASA for this work through grant 
NNG15PP48P to serve as a 
science adviser to the Astro-H project. JHM acknowledges a Herchel Smith Research Fellowship at the University of Cambridge. EJP would like to acknowledge financial support from 
the EPSRC Centre for Doctoral Training in Next Generation Computational 
Modelling grant EP/L015382/1. Figures were generated using Matplotlib \citep{Hunter:2007}. The authors
would also like to thank the reviewer, Maria D{\'{\i}}az Trigo, for many helpful comments which significantly improved the paper.

\bibliographystyle{mnras}
\bibliography{bibliography}
\label{lastpage}

\bsp	

\end{document}